\documentstyle[seceq,supplement,epsf]{ptptex}

\title{
Evaluation of Effective Astrophysical $S$ factor \\
for Non-Resonant Reactions
}

\author{
M. {\sc Ueda},
A. J. {\sc Sargeant}$^{\ast}$,
M. P. {\sc Pato}$^{\ast}$, and
M. S. {\sc Hussein}$^{\ast}$
}

\inst{
Institute of Physics, University of Tsukuba, Tsukuba 305-8571
\\
$^{\ast}$ Nuclear Theory and Elementary Particle Phenomenology Group \\
Instituto de F{\' \i}sica, Universidade de S{\~a}o Paulo \\
Caixa Postal 66318, 05389-970, S{\~a}o Paulo, SP, Brazil
}

\recdate{
}

\abst{
We derived analytic formulas of the effective astrophysical $S$ factor,
$S_{\rm eff}$, for a non-resonant reaction of charged particles
using a Taylor expansion of the astrophysical $S$ factor, $S(E)$,
and a uniform approximation. The formulas will be able to
generate more accurate approximations to $S_{\rm eff}$ than previous ones.
}

\begin{document}

\maketitle


Evaluation of thermonuclear reaction rates is very important for
nucleosynthesis and energy generation in stars. 
The reaction rate at a temperature $T$ is expressed in terms of 
an effective astrophysical 
$S$ factor defined as~\cite{bahcall}
\begin{equation}
S_{\rm eff} = \sqrt{\frac{\tau}{4\pi}}
\frac{e^{\tau}}{E_0} \int_0^{\infty} dE \, S(E)
\exp{\biggl ( -\frac{E}{kT}-2\pi \eta(E) \biggr )} \ ,
\hspace{1cm} \tau = \frac{3E_0}{kT} \ ,
\label{eq0}
\end{equation}
where $k$ is Boltzmann's constant,
$E_0$ the Gamow peak energy, $\eta(E)$ the Sommerfeld parameter,
and $S(E)$ the astrophysical $S$ factor~\cite{clayton}.
In many calculations of the stellar evolution approximations to 
$S_{\rm eff}$ are used. For instance, the approximate expression 
of $S_{\rm eff}$ obtained by the stationary phase approximation~\cite{clayton} 
is made use of for non-resonant reactions of charged particles. 
Since the integrand in Eq. (\ref{eq0}) 
is strongly peaked around $E_0$ (the Gamow peak) so that 
$S_{\rm eff}$ can be almost determined by the contribution from 
the integrand within an energy window of effective width 
$\Delta = 4\sqrt{E_0 kT/3}$~\cite{clayton}.
On the other hand, demand for methods of evaluating $S_{\rm eff}$ more 
accurately increases as theories of stellar evolution are developed. 
Therefore, by using the uniform approximation~\cite{brink} and
Taylor expansion of $S(E)$ around the threashold and the Gamow peak energy,
we made the following two analytic formulas~\cite{mueda} which can generate 
more accurate approximations to $S_{\rm eff}$
\begin{eqnarray}
S_{\rm eff-thd} &=& \sum_{n=0}^{n_M} \frac{1}{n!}S^{(n)}(0) E_0^n
\sum_{k=0}^{k_M} \frac{P_{2k}(n)}{k! (12 \tau)^k} \ , \hspace{1cm}
S^{(n)}(0) = \left . \frac{d^n S}{dE^n} \right \vert_{E=0} 
\label{effthd} \\ 
S_{\rm eff-Gam} &=& \sum_{n=0}^{n_M} \frac{1}{n!}S^{(n)}(E_0) E_0^n
\sum_{r=0}^n (-)^r
\left (
\begin{array}{c}
n \\
r
\end{array}
\right ) \sum_{k=0}^{k_M} \frac{P_{2k}(n-r)}{k! (12 \tau)^k} \ , 
\label{effgam}
\end{eqnarray}
where $P_{2k}(n)$ is the polynomials of $n$ as follwos:
\begin{eqnarray*}
P_0(n) &=& 1 \ , \hspace{1cm}
P_2(n) = 12n^2 + 18n + 5 \ , \\
P_4(n) &=& 144n^2 + 336 n^3 + 84n^2-144n-35 \ , \nonumber \\
P_6(n) &=& 1728n^6 + 4320n^5-4320n^4-13320 n^3 -288 n^2
+6210 n + 665 .
\end{eqnarray*}
The values of $n_M$ and $k_M$ correspond to the numbers
of terms in the Taylor expansion of $S(E)$ and 
the asymptotic expansion of $S_{\rm eff}$ (expansion parameter $1/\tau$),
respectively. The analytic formulas are expected to be useful for 
high temperature enviroments and reactions which have 
strong $E$-dependence around the corresponding Gamow peak energies.
Note that the formulas are valid if $E_0 + \Delta/2$ is 
within the radius of convergence concerning the Taylor expansion of $S(E)$.
\begin{figure}[t]
\centerline{
\begin{minipage}{.45\linewidth}
\epsfxsize=55mm
\epsfbox{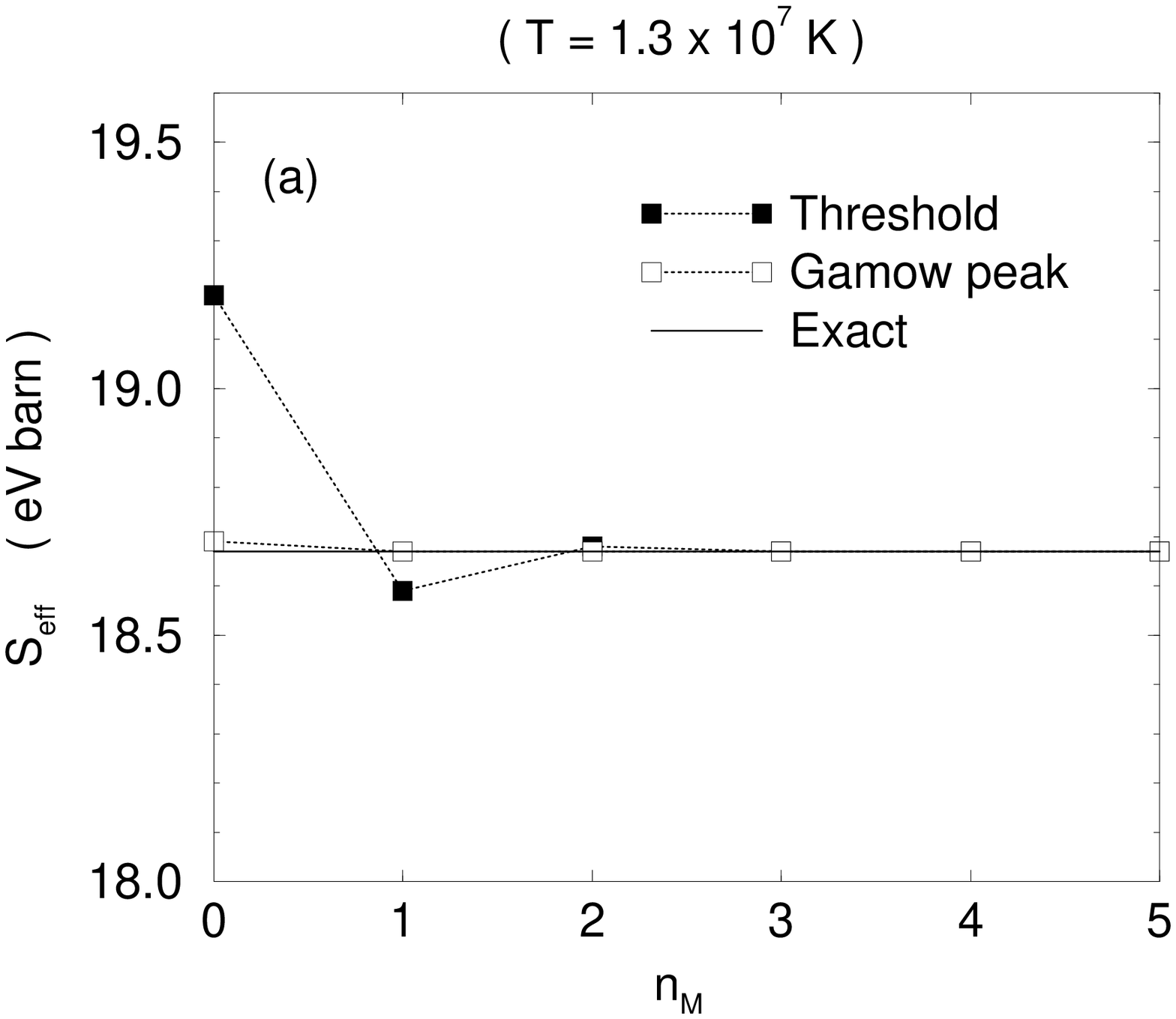}
\end{minipage}
\hspace{10mm}
\begin{minipage}{.45\linewidth}
\epsfxsize=55mm
\epsfbox{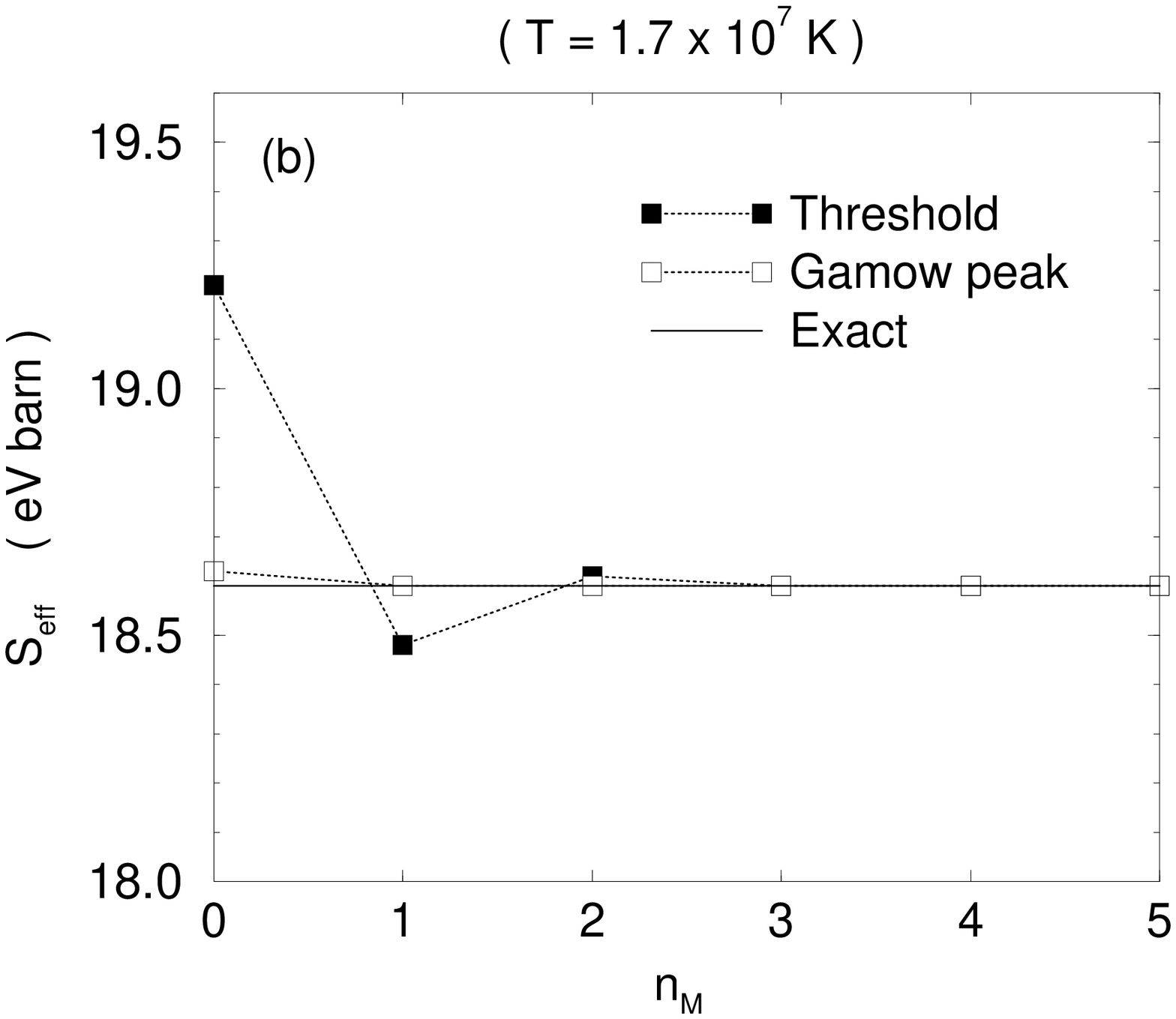}
\end{minipage}
}
\caption{
Approximate values of $S_{\rm eff}$ generated by
$S_{\rm eff-thd}$ (closed squares) and $S_{\rm eff-Gam}$ (open squares) 
together with the exact value of $S_{\rm eff}$ (the solid line)
at (a) $T = 1.3 \times 10^7$ K and (b) $T = 1.7 \times 10^7$ K.
The astrophysical $S$ factor was Taylor-expanded up to $n_M$-th order.}
\label{fig1}
\end{figure}

We applied our formulas to the $^7$Be($p,\gamma$)$^8$B reaction
in the stellar interior of the sun, which is reported to have strong 
$E$-dependence around the corresponding Gamow peak
energy $E_0 \approx 20$ keV~\cite{jen}.
Here we employed the function form of $S(E)$ obtained in 
Ref.~\citen{jen}. Figs. \ref{fig1}-(a) and \ref{fig1}-(b)
show the approximate values of $S_{\rm eff}$ for the reaction at $T = 
1.3 \times 10^7$ K and $T = 1.7 \times 10^7$ K, respectively.
In each figure closed and open squares represent
the approximate values of $S_{\rm eff}$ generated by
Eqs. (\ref{effthd}) and (\ref{effgam}), respectively.
The exact values of $S_{\rm eff}$ is denoted by the solid line.
Note that the two approximate values for a given $n_M$ are not modified by 
inclusion of terms of higher order than $k_M>1$.
Therefore, only the approximate values with $k_M =1$ are shown in the
figures. At both temperatures $S_{\rm eff-thd}$ converges to the exact result 
with $n_M$ = 4 and $k_M$ = 1 while $S_{\rm eff-Gam}$ does with $n_M$ = 2 and 
$k_M$ = 1.  The result can be understood by considering that $S_{\rm eff}$ is
evaluated even quantitatively by $S(E)$ in the energy region 
$E_0-\Delta \leq E \leq E_0+\Delta$. However, it is necessary to confirm
more systematically whether the uniform approximation works well
in approximating $S_{\rm eff}$.

\end{document}